# Resolution recovery on list mode MLEM reconstruction for Dynamic Cardiac SPECT system

Yuemeng Feng[1], William Worstell[2], Matthew Kupinski[3], Lars R Furenlid[3], Hamid Sabet[1]

[1] Department of Radiology, Massachusetts General Hospital & Harvard Medical School, Boston, MA, USA
[2] PicoRad Imaging LLC, Wayland, MA, USA
[3] Center for Gamma-ray Imaging, and the College of Optical Sciences, University of Arizona, Tucson, AZ, USA

E-mail: yfeng16@ and hsabet@mgh.harvard.edu

**Abstract**

The Dynamic Cardiac SPECT (DC-SPECT) system is being developed at the Massachusetts General Hospital, featuring a static cardio focus asymmetrical geometry enabling simultaneous high resolution and high sensitivity imaging. Among 14 design iterations of the DC-SPECT with varying number of detector heads, system sensitivity and system resolution, the current version being fabricated features 10 mm FWHM geometrical resolution (without resolution recovery) and 0.07% sensitivity ; this is 1.5x resolution gain and 7x sensitivity gain compared to a conventional dual head gamma camera. This work presents improvement in imaging resolution by implementing a spatially variant point spread function (SV-PSF) with list mode MLEM reconstruction. A resolution recovery method by PSF deconvolution is validated on list mode MLEM reconstruction for DC-SPECT. A spatial invariant PSF is included as an additional test to show the influence of the accuracy of PSF modelling on the reconstructed image quality. We compare the MLEM reconstruction with and without PSF deconvolution; an analytic model is used for the calculation of system response, and the results are compared to Monte Carlo (MC) based methods. Results show that with PSF modelling applied, the quality of the reconstructed image is improved, and the DC-SPECT system can achieve a 4.5 mm central spatial resolution with average 795 counts/(s*Mbq). The results show substantial improvement over the gold standard GE Discovery 570c performance (spatial resolution 7 mm with an average 460 counts/s*MBq, central resolution 5.8 mm). The impact of PSF deconvolution is significant, and the improvements of the reconstructed image quality is more evident compared to MC simulated system matrix with the same sampling size as in simulation.

## 1. Introduction

We are developing a cardiac dedicated cost-effective stationary SPECT system with large detector coverage (~225 degrees) and an 18 cm diameter field of view (FOV). Balancing design aspects necessitates trade-offs between sensitivity and resolution goals, therefore we have generated 14 design iterations of the DC-SPECT with varying geometry, such as detector pixel size, pinhole size, detector to collimator distances, number of detector heads etc. Previous simulation and modelling results [1-3] demonstrated that our system can achieve a 10.0 mm FWHM system spatial resolution with 0.07% sensitivity. Other than the intrinsic system resolution determined by geometry design, the reconstructed image resolution can also serve as a predictor of the prospective system performance and can be improved by applying iterative algorithms with resolution recovery methods [4-7].

Differing from many current SPECT prototypes, the DC-SPECT has an asymmetric geometry with varying detector to collimator distances, which makes system modelling and reconstruction more challenging in comparison. To ensure high quality of the reconstructed image, the Monte Carlo (MC) simulation-based method is often proposed to accurately model the detector response [8, 9]. However, generating the system matrix for the whole system by MC simulation requires significant computation effort. Analytic calculations, on the other hand, can accelerate the design cycle, but artefacts can propagate to the reconstructed image if accurate detector response modelling is not implemented. Analytical methods can be used to cycle through design iterations, but lack accurately reconstructed images without a MC simulated system matrix. Compensating the mismatch between analytic calculation and the acquired data by involving PSF modelling is an option to balance the computation efforts and the resolution [6, 7, 10, 11], and has shown necessity for improving the image quality [4, 12]. The resolution recovery methods based on PSF are usually proposed for modelling the depth of interaction [13], the finite collimator resolution [14], and the detector blurring as well as the misalignment between the pinhole and aperture [8], and is often modelled as a 2D/3D Gaussian. A novel resolution recovery method based on the pre-reconstruction Fourier transform filtering for MLEM reconstruction is proposed in [15] for small animal SPECT. Our goal is not to compare these different methods; rather we intend to show the impact of PSF deconvolution in the iterative reconstruction for DC-SPECT to demonstrate the necessity of resolution recovery, and to validate an image-based spatial variant PSF (SV-PSF) model.

We first show a comparison between images reconstructed by analytic modelling and MC based methods with low statistics, then investigate an image based PSF deconvolution with analytic modelling in MLEM reconstruction. The PSF is a low pass filter which represents the detector response to a point source and is modelled as a 3D spatial variant Gaussian function with rotation in our study. Results demonstrate the evident improvement of PSF deconvolution in MLEM reconstruction, and show that with appropriate PSF modelling involved, DC-SPECT system can achieve a 4.5 mm central resolution with 0.07% sensitivity. We also prove the computation cost required by modelling the PSF is less than MC simulated system response. Applying the PSF deconvolution with an analytically calculated system matrix in MLEM reconstruction is thus an option to balance the computation cost and the image quality.

## 2. Methods

### 2.1 DC-SPECT system

The DC-SPECT system has previously been implemented in GATE 7.1 and GEANT4 10.1 in [1-3]. The complete system (0.07% sensitivity / 10 mm resolution) involves four rows of pinhole collimator-detector pairs. Each row contains 20 laser-processed [16, 17] CsI:Tl detectors [18], where each 10 mm thick detector is divided into 25x25 converging pixels with size 2x2 mm$^2$. The detailed descriptions for system geometry can be found in [1, 3]. Fig. 1 demonstrates the whole system geometry, with the number 1 to 80 in the figure corresponding to the from first detector to the last. Due to the computation cost, we only considered the first row of 20 heads in this study.

The system resolution in the centre of the FOV can be calculated theoretically by the following factors, as indicated in [19]:

$$R_{system} = \sqrt{R_{coll}^2 + R_{int}^2},$$

$$R_{coll} = d_{eff}\frac{l+b}{l},$$

$$d_{eff} = \sqrt{d * \left(d + 2\mu^{-1}\tan\left(\frac{\alpha}{2}\right)\right)},$$

where $R_{coll}$ is the collimator resolution, $R_{int}$ is the detector intrinsic resolution, $d_{eff}$ is the effective length of the sides of pinhole, l is the length of the collimator, b is the distance from the pinhole centre to the FOV centre, d is the length of sides of the pinhole, $\alpha$ is the acceptance angle of the collimator, $\mu$ is the linear attenuation coefficient of the collimator material, in our case Tungsten ~ 2.5 . Given collimator length 87 mm, distance from the centre of the pinhole to the centre of the FOV 253 mm, pinhole size 2.25 mm and acceptance angle ~42 deg, the theoretical system resolution is estimated as ~ 10 mm. Due to the asymmetrical geometry the resolution and the sensitivity varies within each head. The system sensitivity is calculated by $g = \frac{d_{eff}^2}{16*b^2}$, and the total sensitivity of the complete system with 80 heads is 0.07%.

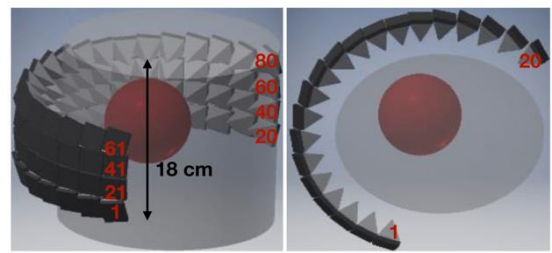

**Fig. 1.** DC-SPECT system, a central cut on transverse plane for the first column of detectors is shown on the right, the scanner covers a total range of 18 cm in transverse plane.

### 2.2 Verification of the uniformity of LOR

It is obvious that with more lines of response (LOR) passing through the voxel, a better resolution can be achieved in the reconstruction. The global detector pixel position and the pixel size are two of the essential factors impacting the distribution of LOR. The imperfect pixel position of one detected gamma photon will add uncertainties to the corresponding LOR. Apart from this, the LOR is also affected by the pinhole shape and the size. The evaluated prototype has 2x2 mm² detector pixel size with depth of interaction 10 mm, and the square-shape pinhole measures 2.25 mm each side. We define the LOR by the centre of the detector pixel and the centre of the pinhole. Figure 2 shows all possible DC-SPECT system LORs intersecting a transverse plane (12 cm each side) in the centre of the FOV, with one and two angles of rotation; 20 heads are considered in the top two figures. Compared to the LORs plotted with 80 heads in the bottom of figure 2, the LORs crossing with 20 heads have less uniformity; however the non-uniformity can be compensated by considering 2 angles of rotation, which corresponds to the current system geometry design, and motivates us to choose the PSF modelling with 2 angles of rotation. Although our DC-SPECT system has an asymmetrical geometry, the uniformity of LORs is proven with 2 angles of rotation and 80 heads; in other words a well-sampling reconstruction volume is assured.

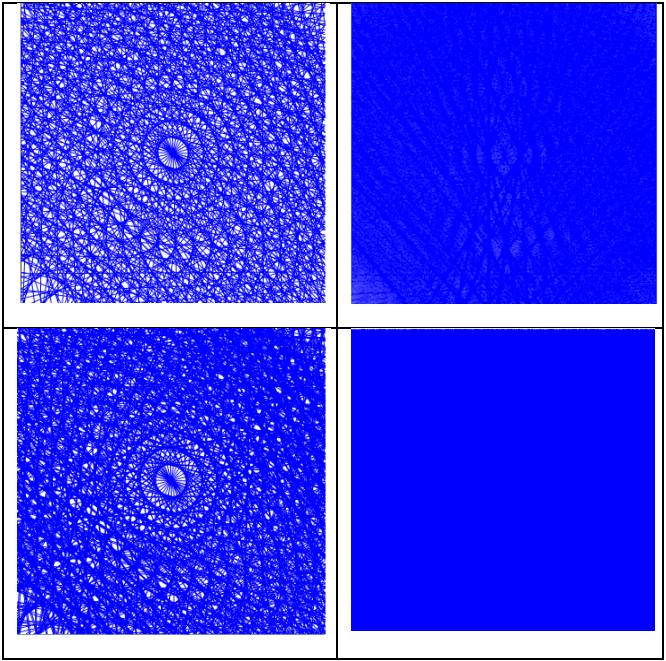

**Fig. 2.** Possible LORs in DC-SPECT system, with one rotation angle $\theta_z$, 20 heads (top left) and 80 heads (bottom left); two rotation angles ($\theta_x$, $\theta_z$) about axis x and z, 20 heads (top right) and 80 heads (bottom right).

## 2.3 List mode MLEM with image based PSF deconvolution.

The list mode MLEM reconstruction algorithm [20] aims to find $\lambda_j$, the mean number of photons emitted in voxel $j$, by iteratively calculating the sequence:

$$\lambda_j^{l+1} = \frac{\lambda_j^l}{S_j} \left( \sum_j \frac{t_{ij}}{\sum_k t_{ik}(\lambda_k^l \otimes k_j)} \otimes k_j \right)$$

where $S_j$ corresponds to the possibility that one event emitted from voxel $j$ is detected by the detector (the so-called sensitivity term) is calculated as $S_j \propto \frac{1}{d^2}$, $d$ is the path length of ray from voxel $j$ to pinhole.

$t_{ij}$ corresponds to one element of system matrix, giving the probability of a photon emitted in voxel $j$ being detected as event $i$; this can be estimated by MC simulation and is analytically calculated in our study by:

$$t_{ij} \propto \frac{1}{d^2}, \text{ if } \frac{\overline{v_iv_j} \cdot \overline{pv_i}}{|\overline{pv_i}|^2} < \sigma,$$

$t_{ij} = 0$ otherwise, $\sigma$ corresponds to one single voxel size in the reconstruction volume, p is the centre of pinhole, $v_j$ is the centre of voxel $j$, and $v_i$ is the detector pixel site of one detected event $i$ referring to the global geometry.

$k_j$ presents the spatially variant PSF and is expressed by a 3D Gaussian with rotation in our study:

$$k_j = k * \exp\left(-\left(\frac{(x'-\mu_x')^2}{\delta_x^2} + \frac{(y'-\mu_y')^2}{\delta_y^2} + \frac{(z'-\mu_z')^2}{\delta_z^2}\right)\right),$$
$$(x', y', z')^T = R * (x, y, z)^T.$$

where (x, y, z) are the voxel coordinates.

R is the 3D rotation matrix in voxel j, containing 2 angles of rotation ($\theta_x$, $\theta_z$):

$$R = \begin{vmatrix} \cos\theta_z & -\sin\theta_z & 0 \\ \cos\theta_x\sin\theta_z & \cos\theta_x\cos\theta_z & -\sin\theta_x \\ \sin\theta_x\sin\theta_z & \sin\theta_x\cos\theta_z & \cos\theta_x \end{vmatrix}.$$

The proposed 3D Gaussian model leads to 9 parameters in total: (k, $\mu_x$, $\mu_y$, $\mu_z$, $\delta_x$, $\delta_y$, $\delta_z$, $\theta_x$, $\theta_z$). The parameters are pre-stored and the PSF for each voxel is calculated on the fly in each iteration of MLEM.

## 2.4 Simulation and SV-PSF modelling

We simulated point-like sources with sampling 1 cm spacing in a 3D Cartesian coordinate system; only the central region of interest (~13 cm in transversal dimension) is evaluated in this study. The coordinates of the source vary from -6.0 cm to 6.0 cm in transverse position, and -1.0 cm to 1.0 cm along each vertical axis, leading to ~500 evaluated positions in total. One point-like source with radius 0.125 mm containing 10 MBq of activity at 140.5 keV was simulated separately in different source centre positions for 10 seconds in air. The 3D Gaussian is then fitted to the reconstructed point after 40 iterations.

The estimated PSF and the spatial variant 3D Gaussian fit on transverse plane, sagittal plane and coronal plane with z=0 are shown in figure 3, each spot corresponds to one point-like source emitted at the location (x, y) cm varying from -6 cm to 6 cm. Results show that the shifts of spatial resolution in different locations in FOV and the reconstructed PSF shape is close to the proposed 3D Gaussian model with rotation. The mean square error (MSE) of the estimated SV-PSF for the total evaluated positions is $1.28 \times 10^{-7}$. We then interpolated the parameters for each voxel.

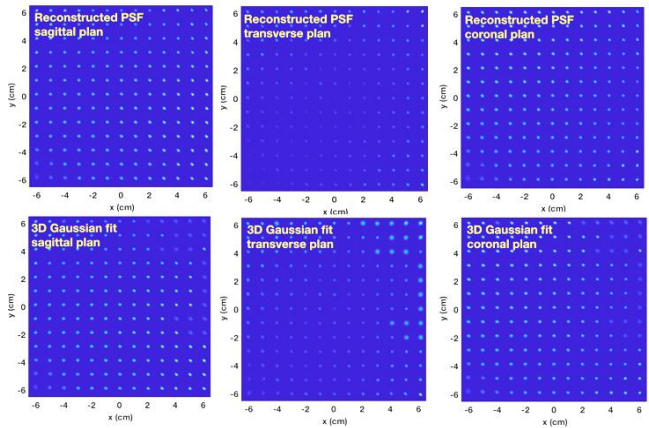

**Fig. 3.** Central slice of PSF reconstructed (top) and estimated (bottom), (x, y) corresponds to the emission position of the point-like sources.

## 3. Results

We reconstructed the modified Derenzo-type phantom shown in figure 4, where each section contains cylinders with length 2.0 cm and diameter 7 mm, 6 mm, 5.5 mm and 4.5 mm. The sources have activity 154 MBq in total, the simulation time is 20 seconds, $2.1 \times 10^6$ events were obtained. We use $2 \times 10^5$ counts for the comparison between system matrix calculation by analytic modelling and by MC based methods. Then $2 \times 10^5$ and $2 \times 10^6$ counts were used separately in paired reconstructions to evaluate the impact of PSF deconvolution. A spatial invariant PSF ($\sigma = 2$) is included in the MLEM as an additional test. The $12 \times 12 \times 2$ cm$^3$ volume is divided into $120 \times 120 \times 20$ voxels (1 mm$^3$ voxels). A post Gaussian smoothing ($\sigma = 1.2$) is added on the central slice of the reconstructed volume within a transverse plane as shown in figure 5, 6 and 7. Figures on the right show profiles along the 2 lines plotted in MLEM reconstruction; line 1 corresponds to resolution 4.5 mm, line 2 corresponds to resolution 5.5 mm. Figure 8 shows the contrast to noise ratio (CNR) and contrast recovery coefficient (CRC) after each iteration for spots with diameter 4.5 mm. The reconstructed image quality metrics are summarized in table 1-2.

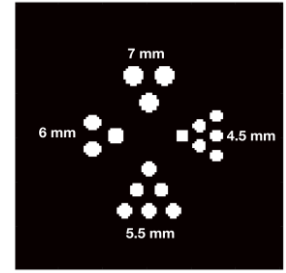

**Fig. 4.** Derenzo-type phantom (ground truth).

TABLE I

COMPARISON BETWEEN ANALYTIC MODELLING AND MC BASED METHODS AT $10^5$ COUNTS, 400 ITERATIONS.

| 4.5 mm, $2 \times 10^5$ counts | Analytic modeling | MC methods |
|---|---|---|
| CNR | 4.2152 | 3.5303 |
| CRC | 8.7284 | 7.3830 |

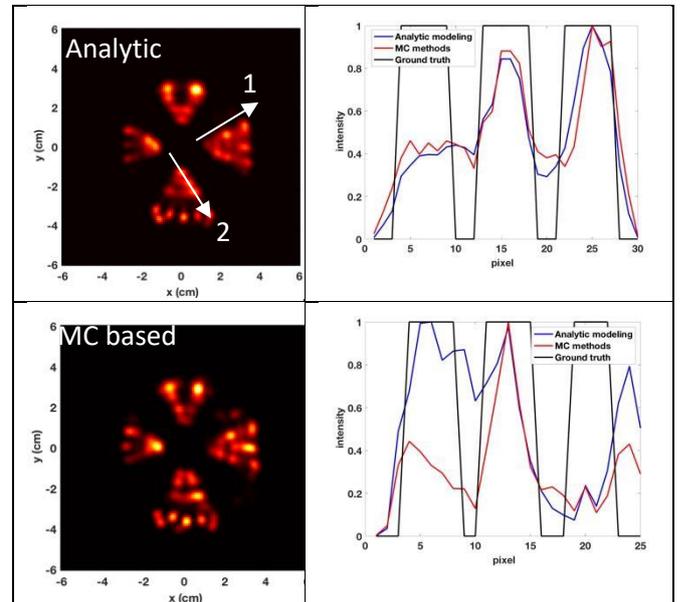

**Fig. 5.** Comparison between analytic modeling and MC based methods.

TABLE II

COMPARISON BETWEEN MLEM WITH ANALYTIC MODELLING, SPATIAL INVARIANT PSF AND SV-PSF, 400 ITERATIONS.

| 4.5 mm, $2 \times 10^5$ counts | MLEM | invariant PSF | SV-PSF |
|---|---|---|---|
| CNR | 4.2152 | 4.4983 | 5.5307 |
| CRC | 8.7284 | 10.0743 | 12.1873 |

| 4.5 mm, $2 \times 10^6$ counts | MLEM | invariant PSF | SV-PSF |
|---|---|---|---|
| CNR | 3.7887 | 4.3587 | 5.3591 |
| CRC | 7.6026 | 9.5934 | 11.7666 |

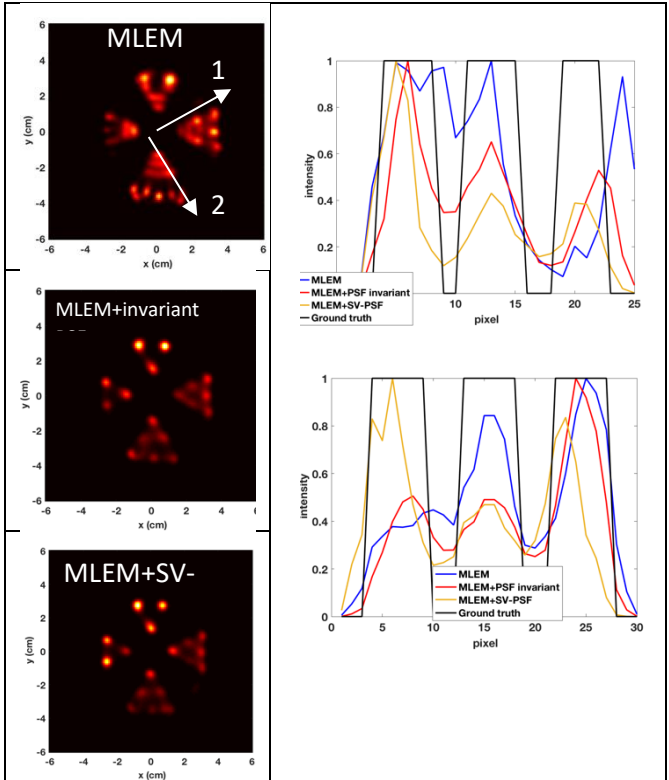

**Fig. 6.** Reconstruction with $2\times10^5$ counts, 400 iterations.

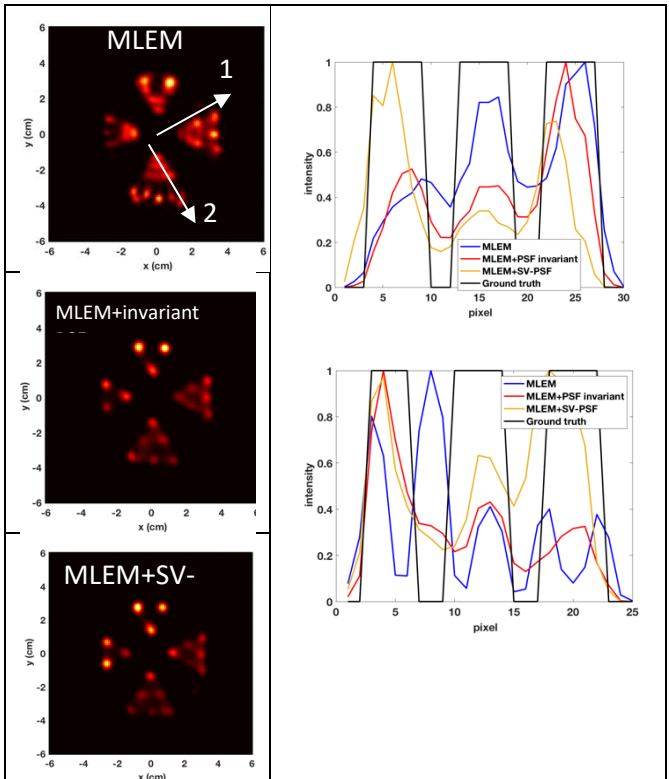

**Fig. 7.** Reconstruction with $2\times10^6$ counts, 400 iterations.

## 4. Discussion

Figure 5 shows that the line profiles of reconstruction by MC based methods is closer to the ground truth for 5.5 mm rods. However, with the current simulation settings, the visual difference between analytic modelling and MC based methods is not obvious. Tab 1 presents the metrics of reconstruction after 400 iterations and shows the advantage of analytic modelling over MC based methods. This may be caused by the large sampling 1 cm, or by the low statistics in the simulation with the acquisition time 20s. With smaller sampling and longer acquisition time, more accurate simulated system response can be expected, and the image quality should be improved. It should be noted that for smaller source spacing, the system matrix generation leads to excessively large computation cost. Currently MC-based method takes ~ 1 hour per source location on a 2.4 GHz CPU. Generating a system matrix with the analytic method is thus nearly instantaneous in comparison with the resources required by the computationally intensive MC-based method.

Figure 3 demonstrates the spatial resolution varying within the FOV, which leads to the necessity of applying a spatial variant PSF for resolution recovery. The rotation of PSF model is caused by the rotation of each detector head in the DC-SPECT geometry and corresponds to the volume sampling shown in figure 2. Nevertheless, results prove that both the SV-PSF and the spatial invariant PSF improve the image quality, and a 4.5 mm resolution can be achieved even with low counts ($2\times10^5$). The reconstruction with SV-PSF generates line profiles closer to the ground truth and provides better CNR and CRC than spatial invariant PSF (figure 8). For SV-PSF the average computation cost measured by the system tick count is 40s per iteration for the selected volume, which is 2 times larger than the invariant PSF (20s); this is caused by the fact that a spatial invariant PSF can be pre-stored and does not require calculation on the fly. Applying the spatial invariant PSF might compensate for the increased computation cost by improving the image quality, especially for clinical applications where lower resolution is sufficient to detect the malignant features of interest.

The MC simulated system matrix had shown no advantage on the image resolution, meanwhile the simulated sampling was 1 cm, so the system matrix measured by MC simulation was not accurate for 1 mm voxel. Comparing the impact of PSF deconvolution with a more accurate MC simulated detector response should be considered in future work.

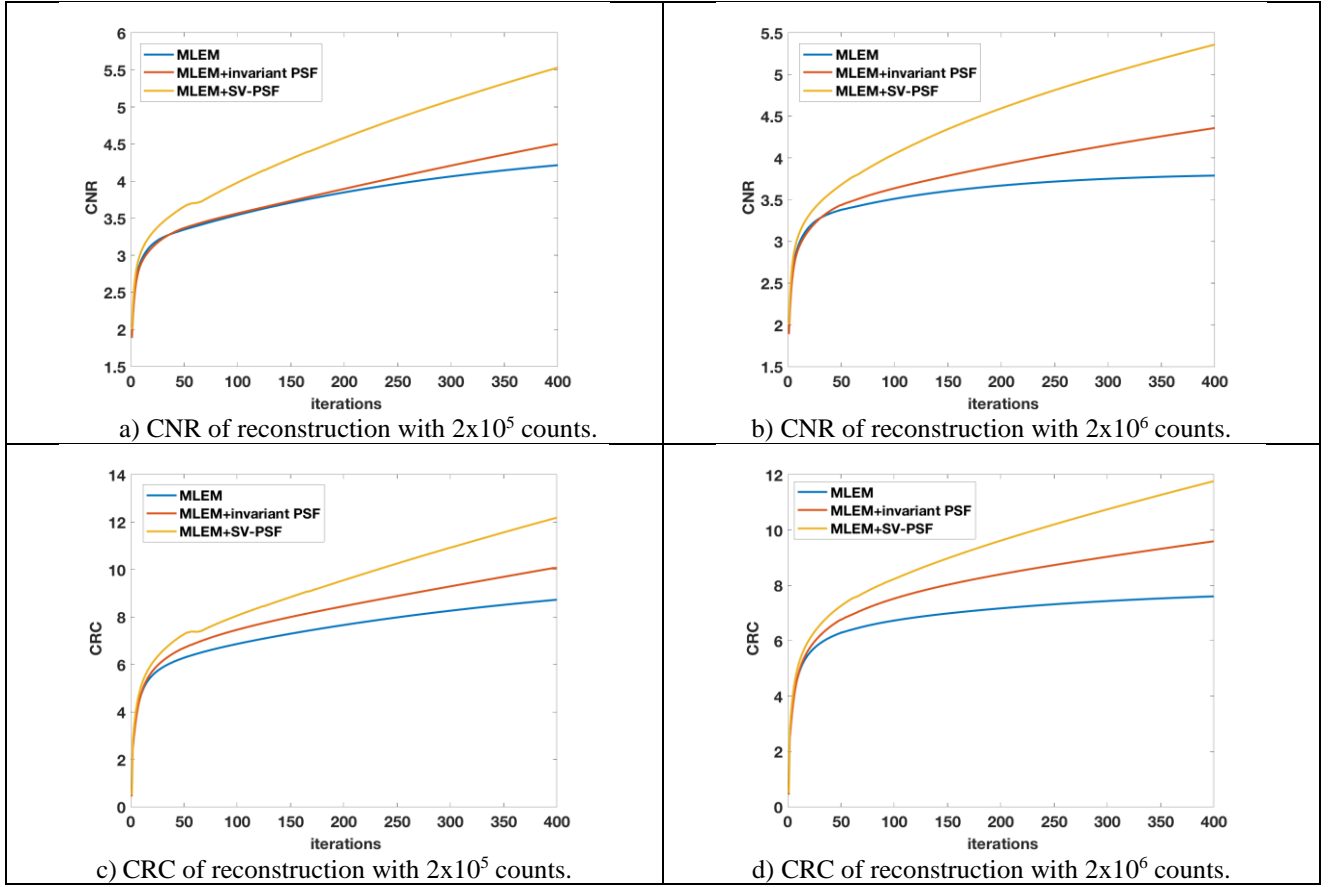

Fig. 8. Metrics of the reconstruction for 4.5 mm rods.

## 4. Conclusion

Results show that with the PSF deconvolution included, 4.5 mm FWHM central spatial resolution can be achieved for DC-SPECT system at 0.07% system sensitivity even with relatively low counts ($2x10^5$), both with the SV-PSF and the spatial invariant PSF. Meanwhile, the spatial resolution varies with source location within the FOV. The impact of the PSF modelling on the border of the FOV will be more significant and will in future be investigated especially when reconstructing an extended source such as the 4D cardiac-torso (XCAT) phantom. With the current simulation settings, reconstruction MC simulated system matrix does not show advantage on improving the image quality and applying the SV-PSF can decrease the artifacts in the reconstruction. We thus propose to apply PSF deconvolution to MLEM with the current analytically-modelled system matrix to improve the reconstruction and to reduce the computation efforts required by MC based methods.

The 4.5 mm central resolution with 0.07% sensitivity is not a definite limit for this system design, to further evaluate the reconstruction performance, attenuation and scattering corrections should also be applied. The complete system contains 80 heads in total, but due to the computation cost we only considered the first row of 20 heads. From the current study, we can only draw the conclusion that with a limited number of detectors, 4.5 mm reconstructed resolution can be expected by list mode MLEM reconstruction with and only with PSF deconvolution. The system performance will be further investigated by including all e 80 heads to have a full coverage in longitudinal direction. We believe that the reconstructed resolution can be improved by involving a larger number of LORs and better sampling over transverse planes.

## Acknowledgements

The authors acknowledge financial support of the NIH Grant No. R01HL145160.